\documentclass[twocolumn,preprintnumbers,prd,nofootinbib,
			 superscriptaddress,nofootinbib]{revtex4}

\usepackage{graphicx, fancybox, color}
\usepackage{amsmath,amssymb,bm}
\usepackage{braket}

\begin{document}

\title{Comment on ``Heavy Quarkonium in Extreme Conditions''}

\author{Masayuki Asakawa}
\email{yuki@phys.sci.osaka-u.ac.jp}
\affiliation{
 Department of Physics, Osaka University, Toyonaka, Osaka 560-0043, Japan
}

\begin{abstract}
In a recent paper (arXiv:1912.02253), Rothkopf claims that
the Bryan method, which is widely used to obtain the solution
in the maximum entropy method and
makes use of the singular value decomposition of a matrix, limits the search
space for the solution. He even presents a counterexample to the Bryan
method. In this comment, we first recapitulate the mathematical
basis of the Bryan method, and reconfirm that it makes use of
no approximations and that it is therefore mathematically rigorous.
In the second part, we explicitly show
that Rothkopf's ``counterexample'' actually does not constitute
a counterexample on the basis of the definition of singular
value decomposition itself.

\end{abstract}

\maketitle

Maximum entropy method (MEM) \cite{jarrel,Asakawa:2000tr}
is one of widely adopted methods
to infer the original images in ill-posed inverse problems.
In nuclear and particle physics, it is, for example, used
to infer the spectral functions from imaginary-time lattice
data with noise. In obtaining the solution, as we show below,
it is required to solve an extremum value problem. The dimension
of the model space where the original image is defined, 
is usually
of ${\mathcal O}(10^3)$. In order to handle this extremum value
problem, an ingenious way by Bryan \cite{bryan} is often utilized.
The method makes use of the singular value decomposition (SVD) of a matrix.
Rothkopf claims that the Bryan method limits the search space
for the extremum problem and that one should search the full
${\mathcal O}(10^3)$-dimensional
vector space \cite{Rothkopf:2019ipj,Rothkopf:2011ef}.
Furthermore, he even presents a ``counterexample''
to the Bryan method (according to him).

In this comment, we first recapitulate the Bryan method and
show that it is mathematically rigorous.
Then, we specify at what point Rothkopf makes a mistake.

In order to explain the Bryan method, let us define the
problem. In the following, we use four discrete variables,
$\omega_l$, $A_l$ ($l=1,2,\cdots,N_\omega$),
$\tau_i$, and $D_{Ai}$ ($i=1,2,\cdots,N$). $\omega_l$,
$A_l$, and $\tau_i$ are discretized
energy, spectral function, and imaginary time, respectively
(see Section 3.4 and Appendix C of Ref. \cite{Asakawa:2000tr}).
$D_{Ai}$ is the discretized imaginary time correlation
function. The imaginary time correlation function $D_A(\tau)$
and the spectral function $A(\omega)$ are related as follows:
\begin{eqnarray}
D_A(\tau) & = & \int_{0}^{+\infty}
\frac{e^{-\tau\omega}+e^{-(\beta-\tau)\omega}}{1-e^{-\beta\omega}}
A(\omega)d\omega \nonumber \\
& \equiv & \int_{0}^{+\infty} K(\tau,\omega)A(\omega) d\omega
\quad (0\leq \tau < \beta).
\label{analytic}
\end{eqnarray}
In the following, $D_{Ai}=D_A(\tau_i)$ is understood.
In (\ref{analytic}), we set the momentum at 0 for simplicity. This,
however, does not affect the following argument.
$N_\omega$ and $N$ are integers of ${{\mathcal O}(10^3)}$
and ${{\mathcal O}(10)}$, respectively.
Let $K_{il}=K(\tau_i,\omega_l)$.
Then, the  problem to solve is expressed as
\begin{equation}
-\alpha \log \left ( \frac{A_l}{m_l} \right )=
 \sum_{i=1}^{N}K_{il}\frac{\partial L}{\partial D_{Ai}},
\label{eqtosolve1}
\end{equation}
where $\alpha> 0$ and $m_l  > 0$ are constants.
Here $L$ is given by
\begin{equation}
L = \frac{1}{2}\sum_{i,j=1}^{N}
 (D(\tau_i)-D_A(\tau_i))C_{ij}^{-1}(D(\tau_j)-D_A(\tau_j)),
\end{equation}
where $D(\tau_i)$ is the data at $\tau_i$ and $C_{ij}$ is the
covariance matrix.
Equation (\ref{eqtosolve1}) is obtained from
the discretized version of the extremum value problem,
\begin{equation}
{\rm max} \left (\alpha S  - L  \right )~~
{\rm with~regard~to~} A(\omega),
\label{problem}
\end{equation}
where $S$ is the Shannon-Jaynes entropy,
\begin{equation}
S = \int _0^{+\infty}
 \left [ A(\omega) -m(\omega) - A(\omega)
 \log \left ( \frac{A(\omega)}{m(\omega)} \right ) \right ]
 d\omega.
\end{equation}
Here $m(\omega) > 0$ is a default model.
By discretizing $S$ as
\begin{equation}
\sum_{l=1}^{N_\omega}
\left [ A_l - m_l - A_l \log \left ( \frac{A_l}{m_l} \right) \right ],
\end{equation}
differentiating
$\alpha S - L$ by $A_j ~(1\le j \le N_\omega)$,
and setting the result to zero, 
Eq. (\ref{eqtosolve1}) is obtained.

Since $A_l \geq 0$ and $m_l > 0$, it is possible to set
\begin{equation}
A_l = m_l \exp a_l \quad (1\leq l \leq N_\omega),
\label{al}
\end{equation}
where $\vec{a} = (a_1,a_2,\cdots,a_{N_\omega})^t~~(a_l \in R)$ is a
general column vector.
Substituting (\ref{al}) into (\ref{eqtosolve1}), one obtains
\begin{equation}
-\alpha \vec{a} = K^t \overrightarrow{\frac{\partial L}{\partial D_A}},
\label{eqtosolve2}
\end{equation}
where $K^t$ is an $N_\omega \times N$ matrix and
$\displaystyle \overrightarrow{\frac{\partial L}{\partial D_A}}$
is an $N$ dimensional column vector.

\newpage
The SVD of $K^t$, which is a real matrix, is defined as $K^t = U\Xi V^t$ \cite{Asakawa:2000tr},
where $U$ is an $N_\omega \times N_\omega$ real
orthogonal matrix satisfying $U^t U = UU^t =1$,
$V$ is an $N\times N$ real orthogonal
matrix satisfying $V^t V = VV^t = 1$, and $\Xi$ is an $N_\omega \times N$
diagonal matrix with positive semi-definite diagonal elements,
$\xi_i~(i= 1,2,\cdots,N)$. $\xi_i$'s can be ordered in such a way
that $\xi_1 \geq \xi_2 \geq \cdots \geq \xi_{N_s} > \xi _{N_{s}+1}
= \cdots = 0$,
where
\begin{equation}
N_s \equiv {\rm rank}~K^t \leq N .
\end{equation}
The explicit form of SVD is
\begin{widetext}
\begin{eqnarray}
K^t & = & U\Xi V^t \nonumber \\
    & = &
\left (
\begin{array}{ccc}
      u_{11} & \cdots & u_{1N_\omega} \\
      \vdots & \ddots & \vdots \\
      u_{N_\omega 1} & \cdots & u_{N_\omega N_\omega}
\end{array}
\right )
\left (
\begin{array}{ccccccc}
     \xi_1 & 0 & & \cdots & & & 0 \\
      0    &   & &        & & &   \\
    \vdots &   & \ddots & \ddots & \ddots & & \vdots \\
           &   & &        & & & 0 \\
      0    &   & & \cdots & & 0 & \xi_N \\
      0    &   & & \cdots & & & 0 \\
      \vdots   &   & &    & & & \vdots \\
      0    &   & & \cdots & & & 0
\end{array}
\right )
\left (
\begin{array}{ccc}
      v_{11} & \cdots & v_{N1} \\
      \vdots & \ddots & \vdots \\
      v_{1N} & \cdots & v_{NN}
\end{array}
\right ) \, .
\label{svd}
\end{eqnarray}
\end{widetext}
Following Bryan \cite{bryan}, we define the $N_s$ dimensional space
spanned by the first $N_s$ columns of $U$ as the ``singular space''.
The bases in this space are
$ \{\vec{u}_1, \vec{u}_2, \cdots, \vec{u}_{N_s} \} $ with
$\vec{u}_i = (u_{1i},u_{2i},\cdots,u_{N_\omega i} )^t$.
From Eqs. (\ref{eqtosolve2}) and (\ref{svd}), one observes
that $\vec{a}$ is in the singular space, whatever
$\displaystyle \overrightarrow{\frac{\partial L}{\partial D_A}}$ is.
This implies that $\vec{a}$ is parametrized only by a set of
$N_s$ parameters $(b_1,b_2,\cdots,b_{N_s})$ as
$\vec{a} = \sum_{i=1}^{N_s}b_i \vec{u}_i$.
Hitherto we have used no approximations.
Each step is based on an elementary theorem
in either analysis or linear algebra. Thus, we have
confirmed that the use of SVD in solving
the extremum value problem (\ref{problem})
preserves mathematical exactitude.
In other words, the Bryan method does not limit the
search space for the solution of the
problem. This is contradictory to Rothkopf's
claim. Therefore, something must be incorrect
in his argument. It is the statement,
\vspace{0.3cm}

{\it ``Now let us choose instead the $N_\tau + 1$st column of $U$
as mock spectrum $\rho$  and compute from it the corresponding
Euclidean data. Then, by construction, this data cannot be
reproduced within one sigma from within the SVD search
space, while it is still possible to reproduce
it in the full search space''} (p.34, ll.16-18 \cite{{Rothkopf:2019ipj}}),
\vspace{0.3cm}

\noindent
that is false. His $\rho$ and $N_\tau $ correspond to our $A$ and $N$,
respectively. Below we explicitly show that this statement
does not hold.

We define
$\vec{D}(\vec{u}_{N+1})=K\vec{u}_{N+1}$.
This is the ``Euclidean data corresponding to $\vec{u}_{N+1}$".
The explicit form of the SVD of $K$ is
\begin{widetext}
\begin{eqnarray}
K & = & V \Xi^t U^t \nonumber \\
    & = &
\left (
\begin{array}{ccc}
      v_{11} & \cdots & v_{1N} \\
      \vdots & \ddots & \vdots \\
      v_{N1} & \cdots & v_{NN}
\end{array}
\right ) \! .
\left (
\begin{array}{cccccccccc}
     \xi_1 & 0 & & \cdots & & & 0 & 0 & \cdots & 0\\
      0    &   & &        & & &   &  & & \\
    \vdots &   & \ddots & \ddots & \ddots &  & \vdots & \vdots & &\vdots \\
           &   & &        & & & 0 & & & \\
      0    &   & & \cdots & & 0 & \xi_N & 0 & \cdots & 0
\end{array}
\right )
\left (
\begin{array}{ccc}
      u_{11} & \cdots & u_{N_\omega 1} \\
      \vdots & \ddots & \vdots \\
      u_{1 N_\omega} & \cdots & u_{N_\omega N_\omega}
\end{array}
\right ) \, .
\label{svdk} 
\end{eqnarray}
\end{widetext}
Since $U^t U = 1$,
\begin{eqnarray}
\vec{D}(\vec{u}_{N+1})=K\vec{u}_{N+1} & = & V\Xi^t U^t \vec{u}_{N+1}
\nonumber \\
& = &V\Xi^t
\left ( 
\begin{array}{c}
      0      \\
      \vdots \\
      1      \\
      \vdots \\
      0
\end{array}
\right )
\leftarrow N+1 \nonumber 
\end{eqnarray}\\[-3ex]
\begin{eqnarray}
~~~~~~~~~~~~~~~~~~~~~~& = & \vec{0} \nonumber \\
& = & K(0\vec{u}_1 + \cdots + 0\vec{u}_{N_s}) \, .
\end{eqnarray}
Note that $0\vec{u}_1 + \cdots + 0\vec{u}_{N_s}$ is in the singular space.
Thus, Rothkopf's statement is proven to be false.
What we have just proven is a part of the following more general fact.
Let $V_{null}$ be the $(N_\omega - N_s)$-dimensional
vector space spanned by $\vec{u}_{N_{s}+1},\vec{u}_{N_{s}+2},\cdots,\vec{u}_{N_\omega}$.
Then,
\begin{equation}
V_{null} = {\rm Ker} (R^{N_\omega} \rightarrow KR^{N_\omega} ). 
\end{equation}
This is the reason why the dimension of
the solution space of the equation,
\begin{equation}
\vec{E} = K\vec{A}
\end{equation}
with $\vec{E}$ being a given $N$-dimensional real vector,
is ${\rm dim}~V_{null} = N_\omega - N_s$ if this equation has at least
one solution \cite{lang}.
This non-uniqueness of the inverse problem was the
very reason to motivate the development of MEM.

In conclusion, Rothkopf's assertion is proven to be invalid.
In Ref.~\cite{Rothkopf:2011ef}, he even found differences between 
the results of the extremum problem (\ref{problem}) with
the SVD method and his ``extended search space'' method.
This fact indicates inaccuracy in his numerical calculations.
Some results presented in
Ref.~\cite{Rothkopf:2019ipj} were obtained with his
method. Hence, Ref.~\cite{Rothkopf:2019ipj} contains
both false mathematical statements and 
unreliable numerical results.

The purpose of this comment is not to claim the superiority
of MEM to other methods, such as analyses on the basis of
pNRQCD \cite{Burnier:2017bod}.
They are complementary to each other if utilized properly.

We thank Masakiyo Kitazawa and P\'eter Petreczky
for discussions on the topics presented in this manuscript.
We acknowledge support from JSPS KAKENHI Grant Numbers JP18K03646.

\end{document}